# The distribution of CO$_2$ on Europa indicates an internal source of carbon

Samantha K. Trumbo[1]* and Michael E. Brown[2]

[1]Cornell Center for Astrophysics and Planetary Science, Cornell University; Ithaca, NY 14853, USA.

[2]Division of Geological and Planetary Sciences, California Institute of Technology; Pasadena, CA, 91125, USA.

*Corresponding author. Email: skt39@cornell.edu.

**Jupiter's moon Europa has a subsurface ocean, the chemistry of which is largely unknown. Carbon dioxide (CO$_2$) has previously been detected on the surface of Europa, but it was not possible to determine whether it originated from subsurface ocean chemistry, was delivered by impacts, or was produced on the surface by radiation processing of impact-delivered material. We map the distribution of CO$_2$ on Europa using observations obtained with the James Webb Space Telescope (JWST) and find a concentration of CO$_2$ within Tara Regio, a recently resurfaced terrain. This indicates the CO$_2$ is derived from an internal carbon source. We propose the CO$_2$ formed in the internal ocean, though we cannot rule out formation on the surface by radiolytic conversion of ocean-derived organics or carbonates.**

Beneath a crust of water ice, Jupiter's moon Europa has an internal ocean of salty liquid water above a rocky seafloor (*1*, *2*), a potentially habitable environment. Assessing the ocean's habitability depends on its chemistry, including the abundances of biologically essential elements, bulk oxidation state, and available chemical energy sources (*3*). The ocean's carbon content is poorly constrained. Previous work has identified CO$_2$ on Europa's geologically young surface via its $\nu_3$ asymmetric stretch fundamental band (*4–6*), however observations possessing the spatial resolution needed to map the CO$_2$ were too limited by noise and artifacts to clearly decipher the source of CO$_2$ (*5*). Attempts to map the CO$_2$ indicated a possible association with dark material, which is potentially endogenic (derived from internal processes), but the results were ambiguous and difficult to interpret (*4*). It has therefore been impossible to distinguish between several possible origins of the CO$_2$: geologically associated CO$_2$ sourced from the ocean; geologically associated CO$_2$ produced radiolytically by the surface bombardment of native organics or carbonates with charged particles; or exogenic (outside Europa) sources, such as delivery by impacts of CO$_2$-rich bodies, or radiolytic conversion from meteorite-delivered carbonaceous material (*7*, *8*).

## CO$_2$ distribution on Europa

We analyze observations of CO$_2$ on Europa obtained with the JWST Near Infrared Spectrograph (NIRSpec) integral field unit (IFU) on 2022 November 23. The spectra have resolving power R~2,700, sufficient to reveal that Europa's CO$_2$ $\nu_3$ band has a double-peaked structure, with minima at 4.249±0.001 and 4.268±0.002 μm (Figure 1). We measure the integrated band area across the entire absorption feature in each spatial pixel, then map the resulting band strengths across the surface *(9)*.

The strongest overall $CO_2$ absorptions are located in Tara Regio (~10ºS, 75ºW; Figure 2A), a roughly 1,800 km diameter area of geologically disrupted resurfaced material (known as chaos terrain) that is among the youngest on the surface *(10, 11)*. The formation of chaos terrain is not well understood; however proposed explanations all involve large-scale disruption of the surface ice via interactions with endogenic material from below, such as upwelling buoyant material, subsurface brines, or lens-shaped formations of liquid meltwater *(12, 13)*. Tara Regio has previously been inferred to contain salty endogenic material *(14, 15)*. Sodium chloride has been detected in the region and interpreted as ocean-derived *(16, 17)*, and it also contributes to the discoloration of Tara Regio relative to the surrounding regions *(16, 18)*.

$CO_2$ is also enhanced within portions of Powys Regio (~0ºS, 150ºW), another large scale chaos terrain, though it is less well-resolved by these observations (Fig. 2A). This region also exhibits generally weaker signatures of endogenic NaCl than does Tara Regio *(15–17)*.

**Potential exogenic sources**

The association between the $CO_2$ band and Europa's geologically young, resurfaced chaos terrain indicates a relationship between the surface $CO_2$ and Europa's internal chemistry. Nevertheless, we consider the possibility of exogenic origins. $CO_2$ has also been detected on two other moons of Jupiter, Ganymede and Callisto *(6, 19–22)*, and throughout Saturn's satellite system *(23–26)*. Unlike on Europa, the $CO_2$ on Ganymede is preferentially associated with dark non-ice material in the oldest, most heavily cratered terrain *(20)*. The $CO_2$ on Callisto is associated with impact craters and has a hemispherical distribution consistent with exogenic processes, thought to be associated with Jupiter's corotating magnetic field *(19, 21)*. Previously suggested explanations for the $CO_2$ on Saturn's icy satellites *(23–26)* include the delivery of $CO_2$-bearing impactors and radiolytic production from externally-derived carbonaceous materials implanted into the surface ice *(23)*.

None of these scenarios would produce the observed relationship between Europa's $CO_2$ and its geologically disrupted large-scale chaos terrain. If the $CO_2$ were externally delivered, we would not expect it to be supplied specifically to terrain formed via endogenic processes. If instead it were produced radiolytically from carbon-bearing meteoritic materials, we expect a distribution reflecting the external implantation, radiation intensity, or both. The leading hemisphere of Europa, where Tara Regio and Powys Regio are located, is thought to receive more meteorite impacts than the trailing hemisphere *(5, 27)*, but there is no reason to expect that input to follow the distribution of chaos terrain. The low latitudes of the leading hemisphere receive higher fluxes of ≥ 20 mega-electronvolt (MeV) energetic electrons, in a longitudinally and latitudinally symmetric lens-like pattern centered on the leading point (0ºN, 90ºW) *(28)*. That does not resemble the highly asymmetric, geologically correlated distribution we observe.

To be stable under Europa's surface temperatures and pressures, $CO_2$ must be trapped within a host material *(5)*. Thus, a more complex alternative to an endogenic explanation could be that the $CO_2$ is created more uniformly across the surface from exogenically implanted material and then somehow trapped more efficiently within chaos terrain. However, there is no evidence for a connection between chaos terrain and minerals known to trap $CO_2$ (including amorphous ice and phyllosilicates (*29, 30*)). The NaCl that is known to be enriched in Tara Regio does not contain the necessary trapping sites in its mineral structure.

We consider whether optical effects, associated with grain size differences and the dark salty material in Tara Regio, could enhance the observed $CO_2$ band depths in this region. Larger ice grain sizes cause photons to encounter more dark particles of non-ice material, which would enhance band strengths if that material were the host for $CO_2$. However, Tara Regio has been suggested to contain smaller ice grains than its surroundings (*31*), which precludes this effect, and an equator-to-pole increase in grain size has been inferred generally across Europa's surface (*5*).

After considering each of these scenarios, we reject the possibility of an exogenic origin of the $CO_2$. Instead, we infer an endogenic origin linked to the geologically young chaos terrain.

**Potential endogenic sources**

In principle, the $CO_2$ may not be the original form of carbon delivered to the surface, but could instead be produced within Tara Regio from the irradiation of emplaced organics (*32, 33*) or carbonate minerals by the ≥ 20 MeV energetic electrons impacting the leading hemisphere (*28*). To investigate these possibilities, we searched the JWST spectra for spectral features around the C-H stretch bands of possible organics (~3.3–3.4 µm) or the ~3.4- and ~3.9-µm bands of carbonates. These signatures of organics have been detected on other icy satellites, such as Enceladus, Iapetus, and Hyperion (*34–36*) (Figure 3A), and carbonates have been observed in the bright spots of Ceres (*37*). We do not detect any such bands in the Europa observations. However, the 3.2–3.6 µm wavelength region in the Europa spectra is complicated by strong bands of water ice, a water-ice reflectance peak at ~3.1 µm, and by a previously identified band of hydrogen peroxide ($H_2O_2$) at 3.5 µm (*38*). If carbonate salts are heavily hydrated, the characteristic bands can be suppressed and hidden by water signatures (*39*).

An endogenic reservoir of organic- or carbonate-bearing precursor material would produce stronger organic or carbonate bands in Tara Regio than in non-chaos terrains. We therefore divide average spectra of Tara Regio by the average of another region to its north, to enhance any organic or carbonate features (Figure 3B). The ratio is dominated by broad differences in the background continuum and near the $H_2O$ 3.1-µm reflectance peak, as well as by the excess $H_2O_2$ already known to be present within Tara Regio (*40*). There are no features attributable to organic material. We therefore exclude the presence of an Enceladus-like organics band at 3.44 µm (Figure 3), with an upper limit of ≲20% of the band strength observed in Enceladus' tiger stripes region (*36*). We also exclude the 3.55 µm band to a similar level; on Enceladus this feature has been interpreted as due to organic material (*36*), possibly methanol ($CH_3OH$) (*41*), or alternatively as $H_2O_2$ (*42*). On Iapetus, the aromatic and aliphatic bands are broad, but the 3.35-µm reflectance peak between them (*34*) would appear in the ratio spectrum (Figure 3B) if this band were present. However, we would not be able to discern similarly broad organics signatures against the stronger $H_2O$ and $H_2O_2$ absorptions if such a reflectance peak were absent (as would occur if dominated by either aromatic or aliphatic compounds).

The ratio spectrum (Fig. 3B) shows a ≲ 1% deviation from the continuum near 4.0 µm, which is coincident with the strongest carbonate absorption band for Ceres' Occator crater. This is consistent with a minor contribution of carbonate minerals in Tara Regio. However, we consider this potential feature to be to too shallow and indistinct in shape to constitute a clear carbonate detection, particularly as it falls at the edge of the gap in JWST spectral coverage. We therefore put an upper limit on any ~3.9 to 4.0-μm carbonate bands equal to the observed depth of this feature (≲ 1%). This is equivalent to a band depth of ≲2.5% of the band observed at Occator crater on Ceres (*37*). No organics or carbonates have been identified in higher-spatial-resolution observations of the same chaos terrain from ground-based observatories (*15*, *40*). Laboratory experiments have investigated the electron irradiation of hydrocarbon and water ice mixtures at Europa-relevant temperatures, finding production of refractory long-chain aliphatic organic material and $CO_2$ (*33*). No equivalent laboratory experiments have been reported for irradiation of carbonates. The JWST data provide no evidence for such additional carbon-bearing material on Europa or preferentially within Tara Regio. We therefore prefer the interpretation that the $CO_2$ we observe on Europa was delivered from an endogenous source, already in the form of $CO_2$.

**Double-peaked $CO_2$ band profiles**

The JWST data resolve two discrete peaks within the $v_3$ band of Europa's $CO_2$ (Fig. 1), which constrains its physical state on the surface. The 4.25-μm peak implies that the associated $CO_2$ exists in a trapped form, causing the fundamental vibration to shift to higher frequencies than the nominal position of ~4.27 μm expected for pure $CO_2$ ice. This phenomenon has previously been invoked to explain the stability and band positions (4.257 to 4.258 μm) of the $CO_2$ bands observed on Ganymede and Callisto, though there is no consensus on the trapping mechanisms and host materials (*19–21*, *30*). The band positions on Europa, Ganymede, and Callisto are all shifted in the opposite direction than expected for $CO_2$ trapped in amorphous water ice (*29*), and do not match the shift expected for $CO_2$ clathrate either, though $CO_2$ in clathrates has two discrete band minima (*43*). At the cryogenic temperatures and near-vacuum pressures on the surfaces of these moons, $CO_2$ adsorbed onto phyllosilicates produces $v_3$ peaks that, in some cases, match those observed on Ganymede and Callisto (*30*), though phyllosilicate minerals have not been identified on their surfaces and we consider them implausible for Europa's salty resurfaced terrain.

We have not identified any laboratory data that both match the wavelength of Europa's 4.25-μm $CO_2$ peak and feature a host material that we consider plausible, based Europa's known surface composition. The substantial difference between the $CO_2$ bands observed on Europa compared to those on Ganymede and Callisto indicates a different trapping mechanism, host material, or both. This is consistent with the apparently different origins of the moons' $CO_2$, which has been associated with older, dark terrain on Ganymede, and with craters and magnetospheric processes on Callisto (*19–21*). The $v_3$ $CO_2$ bands observed on several moons of Saturn are also shifted to higher frequencies, but vary in wavelength by up to 0.014 μm between the different bodies, with Dione and Hyperion providing the closest matches (4.253 μm and 4.252 μm, respectively (*23*)) to Europa's band position, though still not an exact match. The positions of these bands on Dione and Hyperion indicate trapped $CO_2$ associated with other materials or ices, but the specific hosts are unknown (*23*).

The $CO_2$ peak at ~4.27-μm observed on Europa is consistent with the position of 4.267 to 4.268 μm for pure $CO_2$ ice (*43*, *44*), which is too volatile to be stable at Europa's surface conditions (*5*). Either there is another (unknown) trapping mechanism that produces a minimal band shift, or a single trapping mechanism that causes band splitting, perhaps due to distinct trapping sites within a single host material (as in $CO_2$ clathrate, which has a double peak (*43*)). The band depths of both the 4.25-μm and 4.27-μm peaks are strongest in Tara Regio, but their ratio varies across the surface. Within Tara Regio and across the equatorial latitudes, the 4.25-μm peak dominates, while the 4.27-μm peak is stronger across the northern latitudes (Figure 2B), which are colder and more enriched in water ice (*31, 45*). We suggest that low temperatures, ice abundance, or both could be related to the longer-wavelength peak.

**Implications for Europa's atmosphere and ocean**

We interpret the observed association of $CO_2$ with large-scale chaos terrain as indicating emplacement of carbon from the interior. The carbon is probably supplied as $CO_2$ formerly dissolved in the sub-surface ocean, but could also be in the form of other carbon-bearing precursors. Emplacement could have occurred during the formation of such disrupted regions, with potential mechanisms including subsurface upwelling, melt-through, and ice collapse above shallow subsurface liquid (*12, 13*). At the surface, the volatile $CO_2$ must be trapped in one or more other materials, but we expect ongoing sputtering of Europa's surface (*46*) to release some $CO_2$ into Europa's tenuous atmosphere. Photoionization and interactions with Jupiter's magnetosphere would then remove $CO_2$ from the system, as is has been predicted for Callisto (where the expected timescale is ~4 years) (*47*). $CO_2$ that enters the atmosphere—but is not immediately removed—is unlikely to exit the atmosphere, but could instead hop across the surface (as has been simulated for Iapetus (*48*)) until it is either re-trapped or lost. A continual or recent supply of $CO_2$ is therefore required to explain the observed $CO_2$ on the surface, which is consistent with the maximum $CO_2$ band strengths being associated with the youngest terrain.

Our interpretation implies that carbon, a biologically essential element, is present in Europa's subsurface ocean and has reached the surface ice on a geologically recent timescale. If this carbon was delivered as $CO_2$, and if that $CO_2$ is representative of the carbon redox state in the ocean, then a highly reduced ocean chemistry is unlikely (*49*). An oxidized ocean would instead be consistent with the proposed downward delivery (through the ice crust) of radiolytically produced surface oxidants (e.g., $O_2$ and $H_2O_2$) on geologic timescales (*3*). An ocean rich in $CO_2$ would also be consistent with slightly acidic conditions and a metamorphic origin of the ocean (*50*).

**Acknowledgments:** Based on observations made with the NASA/ESA/CSA James Webb Space Telescope. The data were obtained from the Mikulski Archive for Space Telescopes at the Space Telescope Science Institute, which is operated by the Association of Universities for Research in Astronomy, Inc., under NASA contract NAS 5-03127 for JWST. These observations are associated with program #1250. The authors gratefully acknowledge Heidi Hammel and the JWST GTO #1250 team led by Geronimo Villanueva for developing the Europa observing program and providing it to the community with no exclusive-access period.

**Funding:** SKT is funded by a Heising-Simons Foundation 51 Pegasi b postdoctoral fellowship 2021-2663

**Author contributions:** SKT performed the analysis, led the interpretation, and wrote the manuscript. MEB reduced the JWST data, contributed to the interpretation, and aided in manuscript writing.

**Competing interests:** The Authors declare that they have no competing interests.

**Data and materials availability:** The JWST data are available at the NASA Mikulski Archive for Space Telescopes under program ID 1250 (https://mast.stsci.edu/portal/Mashup/Clients/Mast/Portal.html). Data from program ID 1538 were used for calibration *(52)*. Data reduction software is archived at Zenodo *(53)*.

**Supplementary Materials**

Materials and Methods

References (*54–57*)

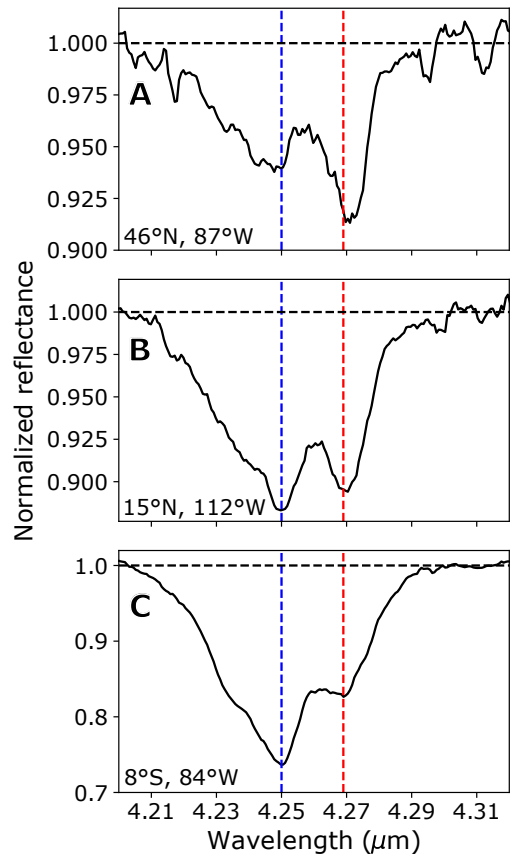

**Fig. 1. Spectra of the $CO_2$ $\nu_3$ band on Europa.** Continuum-normalized spectra are shown by the solid black curves, while the dashed black lines indicate the continuum-level. The two band minima are marked by vertical blue and red dashed lines at 4.25 μm and 4.269 μm, respectively. **(A)** Spectrum corresponding to a spatial pixel centered at 46ºN, 87ºW, in which the 4.27-μm peak dominates within the $\nu_3$ band. **(B)** Spectrum from a pixel centered at 15ºN, 112ºW, in which the 4.25-μm and 4.27-μm peaks are close in depth. **(C)** Spectrum corresponding to a pixel

centered at 8ºS, 84ºW within Tara Regio, where the 4.25-μm peak is stronger than the 4.27-μm peak. The spectra shown in (A), (B), and (C) have each been smoothed using a 5-point moving average, and their corresponding pixel locations are outlined in white in Figure 2B.

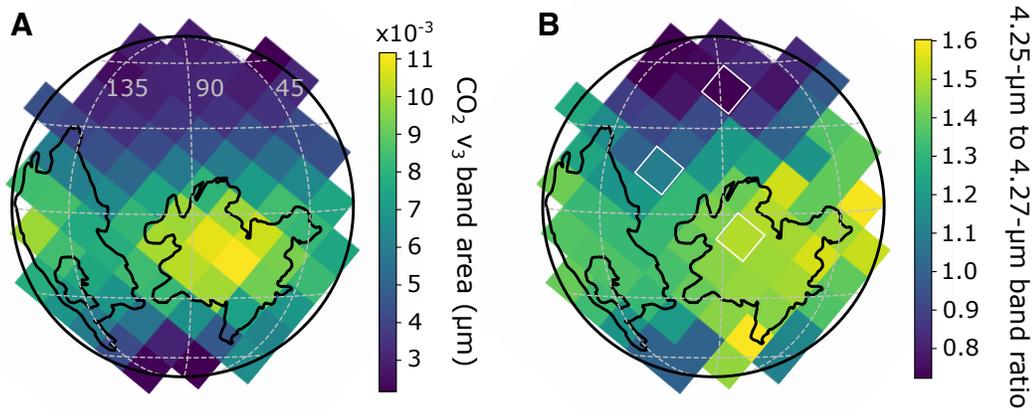

**Fig. 2. Maps of the $CO_2$ $\nu_3$ band area and the ratio between its two peaks.** The disk of Europa is indicated by the black circles in both panels and centered at 2.7°N and 93°W. Gray dashed lines indicate the 45°W, 90°W, and 135°W meridians and the 60°S, 30°S, 0°N, 30°N, and 60°N parallels. The large-scale chaos regions Tara Regio (~10ºS, 75ºW) and Powys Regio (~0ºS, 150ºW) are outlined in black within the disk using the boundaries as defined by (*11*). **(A)** Mapped band area (indicated by the color bar) of the entire $\nu_3$ $CO_2$ band. The strongest absorption occurs in Tara Regio, the region of disrupted terrain centered to the right of the 90°W meridian. $CO_2$ also appears concentrated in parts of the chaos region Powys Regio, which is shown on the left portion of the disk. **(B)** The mapped ratio between the 4.25-μm and 4.27-μm peaks within the $CO_2$ band. The spatial pixels outlined in white mark the locations of the spectra

shown in Figure 1. The 4.25-μm peak is the stronger of the two within the chaos terrain and at low latitudes, while the 4.27-μm peak is stronger in the colder, more ice-rich northern latitudes.

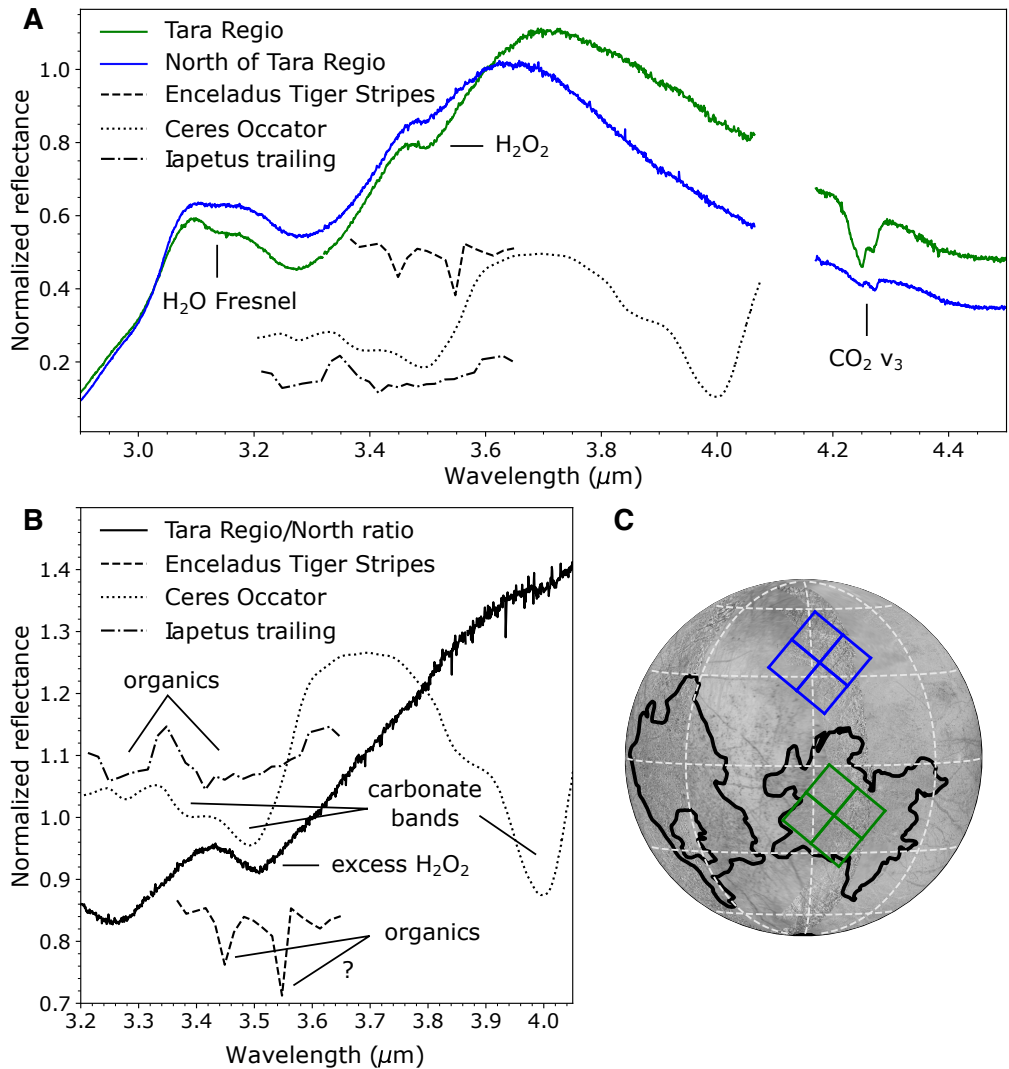

**Fig. 3 Comparison between spectra of Europa and of other Solar System bodies. (A)** Spectra (average of 4 spatial pixels) of Tara Regio (green) and a location north of Tara Regio (blue). Labels indicate the $\nu_3$ band of $CO_2$, the 3.5-μm band of hydrogen peroxide ($H_2O_2$), and the 3.1-μm reflectance peak of water ice. Also shown for comparison are a spectrum of Occator crater on Ceres (dotted black line), which has absorption due to carbonate minerals (*37*); a spectral ratio of the tiger stripes region relative to nearby regions on Enceladus (dashed black line) (*36*), which exhibits absorptions interpreted as organics; and a continuum-removed spectrum of the trailing hemisphere of Iapetus (dash-dotted black line), which shows bands at ~3.3 μm and ~3.45 μm due to aromatic and aliphatic organics, respectively (*34*). All spectra have been normalized at 3.6 μm and the Ceres, Enceladus, and Iapetus spectra are offset for display by 0.55, 0.5, and 0.78 units, respectively. **(B)** The ratio (solid black line) between the spectra of Tara Regio and the region to its north from panel A. The same spectra of Enceladus, Iapetus, and Ceres as in panel A are included for comparison, with additional bands labelled. We attribute the 3.5-μm band in the ratio spectrum to excess $H_2O_2$, which has previously been seen to be more abundant in Tara Regio (*40*). No other absorption bands are evident in the ratio spectrum, except for a ≲ 1% deviation from the continuum near 4.0 μm, which is too shallow to indicate abundant carbonates and immediately borders the gap in JWST spectral coverage. **(C)** Optical imaging mosaic of Europa projected to the JWST observing geometry. Overlain are the boundaries of the chaos terrain (black outlines *(11)*) and the locations of the NIRSpec pixels we used to produce the average spectra of Tara Regio (green squares) and the terrain to its north (blue squares) in panel A. Dashed white lines mark the 45°W, 90°W, and 135°W meridians and the 60°S, 30°S, 0°N, 30°N, and 60°N parallels. The background Europa image is the USGS Voyager – Galileo global mosaic *(51)*.

# Supplementary Materials for

The distribution of $CO_2$ on Europa indicates an internal source of carbon

Samantha K. Trumbo and Michael E. Brown

Correspondence to: skt39@cornell.edu

**This PDF file includes:**

Materials and Methods
References 54–57

**Materials and Methods**

Data reduction

Europa was observed with JWST NIRSpec *(54)* at approximately 08:18 Universal Time on 2022 November 23 as part of the Cycle 1 Guaranteed Time Observations program #1250. The observations were obtained with the integral field unit (IFU), F290LP filter, and G395H high-resolution grating, which provide a spectrum from 2.9 to 5.2 μm with a resolving power of ~2,700. A two-point dither pattern was used with an effective exposure time of 268.4 seconds per dither. The observations were centered on Europa's leading hemisphere at a sub-observer latitude of 2.7°N and sub-observer longitude of 93°W.

We obtained the observations from the JWST archive. At the time of our analysis, the standard JWST pipeline (version #1.8.5) *(55)* could not correctly combine multiple dithers of moving targets or account for 1/f noise (frequency-dependent noise, where f is frequency) from electronic drifts during readout of the detector. During the `cube_build` step of the reduction, which interpolates data from the one-dimensional slits to produce two-dimensional images at every wavelength slice, the standard pipeline also introduces artificial wave-like systematics to the IFU spectra. We therefore implemented a custom pipeline reduction designed to mitigate these issues *(53)*.

We start with the Level 2 two-dimensional calibrated "rate" files, which contain 1/f noise, and use the detector pixels not exposed to sky to isolate the signal modulations from this noise. We identify these unilluminated pixels from the Level 2 "cal" files and then subtract a local median of unilluminated pixels within ±150 rows from each pixel in the "rate" image. We then feed these modified "rate" files into the remainder of the Level 2 pipeline, but modify the `cube_build` step to circumvent most of the wave-like systematics associated with the standard interpolation. Instead of the default `cube_build`, which builds the image cubes in the sky coordinate frame, we have the pipeline assemble the cubes with one axis parallel to the slits (using the "ifualign" geometry), so that minimal interpolation is applied. Rows in our resulting data cubes correspond to individual slits. The resulting spatial pixels have an angular size of 0.1″ × 0.1″, which matches the pixel scale of the default pipeline cubes. For each of the two dither positions, the spectrum is dispersed onto two individual detectors ("NRS1" and "NRS2"). We reduce each dither/detector combination separately.

Next, we cross-correlate each dithered pair to determine their pixel offsets, then apply shifts to align and combine the cubes of separate dithers into a single image cube per detector. We assume integer pixel offsets, which may result in degraded spatial resolution by up to half of a pixel. In combining the dithers, we normalize the two spectra at each spatial pixel to the same median value and then take the median at each wavelength.

To convert the Europa spectra to reflectance, we use JWST observations of the G0V star GSPC P330-E (Program 1538 *(52)*) as a solar analog spectrum. These stellar data were taken with identical instrument settings, so allow us to divide out the spectral response of the instrument, which imparts its own wavelength-dependent distortions. We reduce the stellar data in the same way as the Europa observations, except we do not combine individual dithered cubes. Instead, we perform point-spread-function (PSF) fitting to extract a spectrum from each dither position and take a median of all of the extracted spectra as our final stellar spectrum.

Most of the narrow spectral features visible in the Europa spectra are solar absorption features, which allow us to correct the (otherwise unknown) error in the wavelength calibration by comparing to a high-spectral resolution solar Kurucz model *(56)*. We smooth and bin the model spectrum to the JWST spectral resolution and sampling, accounting for the Doppler shifts resulting from Europa's motion relative to the Sun and JWST, to put the model in the observer reference frame, then cross-correlate the result with the Europa data. We find the wavelength error to be < 0.33 pixels or $1.8 \times 10^{-4}$ µm, which is negligible, but we nevertheless apply this correction to the Europa wavelength scale. We then divide by the measured GSPC P330-E stellar spectrum, which we also shift by a sub-pixel amount (0.67 pixels or $3.5 \times 10^{-4}$ µm determined via cross-correlation) to account for the star's velocity and unknown wavelength error.

We present these spectra in the JWST rest frame in which they were observed. The wavelength-dependent Doppler shifts from Europa's rest frame due to its motion relative to the observer (~$1 \times 10^{-4}$ µm across the region of interest) are negligibly small compared to the shifts and widths of the $CO_2$ features we analyze and within the uncertainties on the measured $CO_2$ band positions.

Our approach avoids the majority of the wave-like systematics induced by the standard pipeline, but some pixels close to Europa's limb are still unavoidably affected. This is because the instrument undersampling of the point spread function inherently results in flux oscillations during slit rectification, which can particularly influence the limb pixels, where Europa's signal is changing rapidly. We exclude such pixels from our analysis.

As a final step in the data reduction, we determine the geographic coordinates of each spatial pixel within the Europa image cubes. We find the center location in each cube that maximizes the total signal within a disk of Europa's angular diameter (0.98″ at the time of observation) and take this to be the sub-observer point. We then calculate the coordinates of each spatial pixel using the relevant position angles of the observations and the corresponding target geometry information as obtained from JPL Horizons *(57)*.

Band strength and position measurements

To measure the strength of the $CO_2$ absorption and the positions and ratios of the $CO_2$ minima within each spatial pixel, we perform continuum fitting and removal on each associated spectrum. We fit a straight line from 4.19 to 4.33 µm, excluding the 4.21 to 4.3 µm region across the absorption band, divide out this defined continuum, and then integrate the residual absorption to obtain the band area (or equivalent width) of the entire $CO_2$ feature in each spatial pixel. To define the individual locations of the two peaks visible within Europa's $CO_2$ band, we fit separate Gaussian functions to the 0.015 µm region surrounding each minimum in the continuum-removed spectra. We measure the relative strengths of the two peaks by taking the ratio of the resultant band depths at the minima of each Gaussian. We find mean band positions of 4.249±0.001 and 4.268±0.002 µm, where the uncertainty is the standard deviation across all pixels.

Constraints on organic and carbonate features

We assess the Europa spectra for organic or carbonate absorptions and compare with analogous detections on other bodies in the Solar System. Because the spectra show no organic features in the 3.2 to 3.6-µm region that are discernible against the background water-ice and $H_2O_2$ signatures, we place limits on the possible strength of undetected absorption features analogous to the organic bands observed on Enceladus *(36, 41)*. We attempted to fit the relevant region of the spectrum, divide out the fitted continuum, then assess the scale of the residuals. However, we find that the 3.2 to 3.6-µm range is not well-fit by a low-order polynomial nor by a water-ice model. We ascribe this to Europa's surface containing salts and other, unidentified components, for which suitable models are not available. Higher-order polynomials or cubic splines can fit the continuum, but such functions may be over-fitting the weak bands we are seeking. We therefore adopt an alternative approach by multiplying the Europa spectra by the continuum-removed Enceladus bands at varying fractions of their strengths, to assess the level at which the absorptions become distinguishable from the continuum. Using an average spectrum from the four pixels centered within Tara Regio, we find that Enceladus-like organic bands become discernible from the continuum at approximately 20% of their strength at the Enceladus Tiger Stripes (*36*). This is a constraint on the overall feature strength that could be obscured by Europa's much-stronger water-ice and $H_2O_2$ absorptions, not on the potential abundance of organics.

To investigate the possibility of endogenic organic- or carbonate-bearing material preferentially associated with Tara Regio, we take the same approach using the ratio between the average Tara Regio spectrum and a four-pixel average from north of Tara Regio, where we see weaker $CO_2$ bands (Fig. 3A). We exclude the observed Enceladus organic bands to an estimated 20% of the strength seen within the Tiger Stripes (*36*). The ratio spectrum exhibits a notch near 4 μm (Fig. 3B). We assess its deviation from the continuum level to constrain the strength of any carbonate feature in this region. We exclude the presence of a strong 4-μm carbonate band, like that observed within Ceres' Occator crater (*37*) to ≲2.5% of the depth at Occator. The 4-μm feature in Tara Regio is too indistinct to constitute a detection.